\documentclass[aps,prx,twocolumn,superscriptaddress,a4paper,english,longbibliography]{revtex4-1}

\usepackage{times}
\usepackage{color}
\usepackage[colorlinks=true,urlcolor=blue,citecolor=blue]{hyperref}
\usepackage{url}
\usepackage{breakurl}
\usepackage{soul}
\usepackage{graphicx}
\usepackage{amsmath}
\usepackage{subfigure}
\usepackage{xcolor}
\usepackage{amssymb}
\usepackage{latexsym}
\usepackage{hyperref}
\DeclareGraphicsExtensions{.png,.eps}
\usepackage{float}
\usepackage{ulem}
\usepackage{appendix}
\usepackage{etoolbox}

\newcommand {\R}{\textcolor {black}}
\newcommand {\B}{\textcolor {black}}

\usepackage{xcolor}
\usepackage[urlcolor=blue]{hyperref}      
\hypersetup{
    colorlinks = true,                    
    citecolor = {blue},
    linkcolor = {purple},
           }
\begin{document}	
	
\title{Boundary-induced singularity in strongly-correlated quantum systems at finite temperature}

\author{Ding-Zu Wang}
\affiliation{School of Physics, Beihang University,100191,Beijing, China}

\author{Guo-Feng Zhang}
\email[Corresponding author: ]{gf1978zhang@buaa.edu.cn}
\affiliation{School of Physics, Beihang University,100191,Beijing, China}

\author{Maciej Lewenstein}
\email[Corresponding author: ]{maciej.lewenstein@icfo.eu}
\affiliation{ICFO - Institut de Ciencies Fotoniques, The Barcelona Institute of Science and Technology, Av. Carl Friedrich Gauss 3, 08860 Castelldefels (Barcelona), Spain}
\affiliation{ICREA, Pg. Llu\'is Companys 23, 08010 Barcelona, Spain}

\author{Shi-Ju Ran}
\email[Corresponding author: ]{sjran@cnu.edu.cn}
\affiliation{Department of Physics, Capital Normal University, Beijing 100048, China}
\date{\today}

\begin{abstract}
	Exploring the bulk-boundary correspondences and the boundary-induced phenomena in the strongly-correlated quantum systems belongs to the most fundamental topics of condensed matter physics. In this work, we study the bulk-boundary competition in a simulative Hamiltonian, with which the thermodynamic properties of the infinite-size translationally-invariant system can be optimally mimicked. The simulative Hamiltonian is constructed by introducing local interactions on the boundaries, coined as the entanglement-bath Hamiltonian (EBH) that is analogous to the heat bath. The terms within the EBH are variationally determined by a thermal tensor network method, with coefficients varying with the temperature of the infinite-size system. By treating the temperature as an adjustable hyper-parameter of the EBH, we identify a discontinuity point of the coefficients, dubbed as the ``boundary quench point'' (BQP), whose physical implication is to distinguish the point, below which the thermal fluctuations from the boundaries to the bulk become insignificant. Fruitful phenomena are revealed when considering the simulative Hamiltonian, with the EBH featuring its own hyper-parameter, under the canonical ensembles at different temperatures. Specifically, a discontinuity in bulk entropy at the BQP is observed. The exotic entropic distribution, the relations between the symmetries of Hamiltonian and BQP, and the impacts from the entanglement-bath dimension are also explored. Our results show that such a singularity differs from those in the conventional thermodynamic phase transition points that normally fall into the Landau-Ginzburg paradigm. Our work provides the opportunities on exploring the exotic phenomena induced by the competition between the bulk and boundaries.

\end{abstract}
\maketitle

\section{Introduction}
The boundary effects in quantum many-body systems have witnessed a surge of interest in a wide range of issues from quantum entanglement~\cite{PhysRevLett.96.100603, 2009JPhA...42X4009A, PhysRevD.93.084021, PhysRevB.100.235112} \B{and phase transitions}~\cite{PhysRevE.91.042123, 2015JSMTE..11..015C, PhysRevLett.118.040601, PhysRevB.98.085111} \B{to the topological superconductors}~\cite{2006Sci...314.1757B, RevModPhys.82.3045, PhysRevLett.116.133903, PhysRevB.103.075126, RevModPhys.83.1057, PhysRevB.90.220511, PhysRevB.103.144512}. Impressive progress has been achieved on the artificial platform for probing the interesting but complex strongly-correlated systems~\cite{RevModPhys.86.153, PhysRevB.98.085111, PhysRevB.99.205132, PRXQuantum.3.020304} \B{by utilizing the boundary effects}. Such schemes about boundary effects also shed light on studying the quantum dynamics, e.g., anomalous transport processes~\cite{PhysRevB.80.052301, PhysRevLett.108.180601} and Markovian closure~\cite{PhysRevLett.129.140604}. 

An inspiring achievement in \B{unveiling the non-trivial physics in the bulk of} the strongly-correlated systems concerns the entanglement Hamiltonian 
(EH)~\cite{1975JMP....16..985B, 1976JMP....17..303B, PhysRevLett.121.200602, PhysRevB.99.235109, PhysRevResearch.3.013217}. The EH is defined as the negative logarithm of the reduced density matrix which describes the bulk as a thermal state. It offers insights into intriguing problems such as the subsystem thermalization~\cite{2005JSMTE..04..010C, RevModPhys.83.863, 2016JSMTE..12.3103C, 2018JSMTE..11.3103W} and the simulation of quantum many-body thermodynamics~\cite{2021NatPh..17..936K}. How the EH could be applied to investigate the many-body thermodynamics with non-trivial boundaries is still elusive.

In this work, we investigate the thermodynamic properties induced by the competition between the bulk and the ``infinite'' boundary defined by the entanglement-bath Hamiltonian (EBH)~\cite{PhysRevE.93.053310, PhysRevB.96.155120, PhysRevB.105.155155}. The EBH was originally proposed for constructing a finite-size \R{simulative Hamiltonian} to access the properties of the infinite-size translationally-invariant (TI) system, where the finite-size effects~\cite{2018QS&T....3d4006S} are effectively compressed \R{by the EBH located on the boundaries}. In other words, the terms in the EBH are variationally determined by tensor network (TN) method so that the bulk entanglement Hamiltonian of the finite-size simulator mimics that of the infinite-size system. 

\R{When considering to simulate the thermodynamics at a given temperature (denoted as $T'$), our results show that the terms in the EBH are determined by the local Hamiltonian of the infinite-size TI model, and the optimal values of the coefficients vary with $T'$. We uncover a discontinuous point} of the coefficients with respect to $T'$, which we name as the ``boundary quench point'' (BQP). 

\R{The physical implication of BQP concerns the thermal fluctuations ``felt'' by the bulk from the environment. It is natural that the thermal fluctuations decrease as $T'$ decreases, and eventually the thermal effects from environment to bulk become identical to those of the ground state (i.e., $T'=0$). Our results suggest that such a decrease is not smooth. The BQP is the singular point, below which the thermal fluctuations from environments to bulk become insignificant, and the bulk is affected by the environments like those of the ground state.}

\R{Our main goal is the exotic thermodynamics by regarding $T'$ not as a physical temperature but as a hyper-parameter that controls the coefficients of the EBH. Below, we construct the simulative Hamiltonian by following the standard way explained above, and study the thermodynamics by constructing the canonical ensemble from the simulative Hamiltonian at the temperature $T$. The critical difference here is that we adjust $T'$ independently} from the physical temperature $T$ in the canonical distribution. 

\R{Equivalently speaking, there are two main steps in our simulation. The first step is to obtain the finite-size simulative Hamiltonian that mimics the thermodynamics of the infinite-size TI system at the temperature $T'$ (namely to take $T=T'$). The second step is to construct a simulative Hamiltonian with the $T'$-parameterized EBH on its boundaries and study the thermodynamics of such a simulative Hamiltonian at the temperature $T$. Since the EBH already contains certain temperature information $T'$, competitions between the bulk and boundary will emerge for $T \neq T'$. Our results of the thermodynamic entropy show that the EBH in such cases can act like the heat bath, even though the physics are given by a canonical ensemble. The bulk entropy is shown to be discontinues with respect to $T'$ at the BQP.}

\R{We also investigate the relations between the symmetries of Hamiltonian and the position that BQP appears} (dubbed as $T'_{Q}$). For the spin-1/2 and spin-1 XXZ chains, the BQP disappears at the Heisenberg point where the Hamiltonian satisfies the SU(2) symmetry, and emerges when reducing to the XY-type or Ising-type Hamiltonian dominated by U(1) or Z$_{2}$ symmetry. The impacts from \R{the bath dimension $\chi$ (referring to the local degrees of freedom of each bath site, which is adjustable)} are also investigated. For a XY-type Hamiltonian, the BQP moves to a low temperature $T'_{Q}$ as $\chi$ increases, and the discontinuity tends to disappear in the large $\chi$ limit. For an Ising-type Hamiltonian, the BQP converges to a finite $T'$ and the discontinuity of entropy at the BQP is shown to be robust. \R{Our work suggests that the singularity at the BQP} differs from those in the conventional thermodynamic phase transitions that normally fall into the Landau-Ginzburg paradigm. 

\section{Entanglement-Bath Hamiltonian}

\begin{figure}[tbp]
	\centering
	\includegraphics[angle=0,width=1\linewidth]{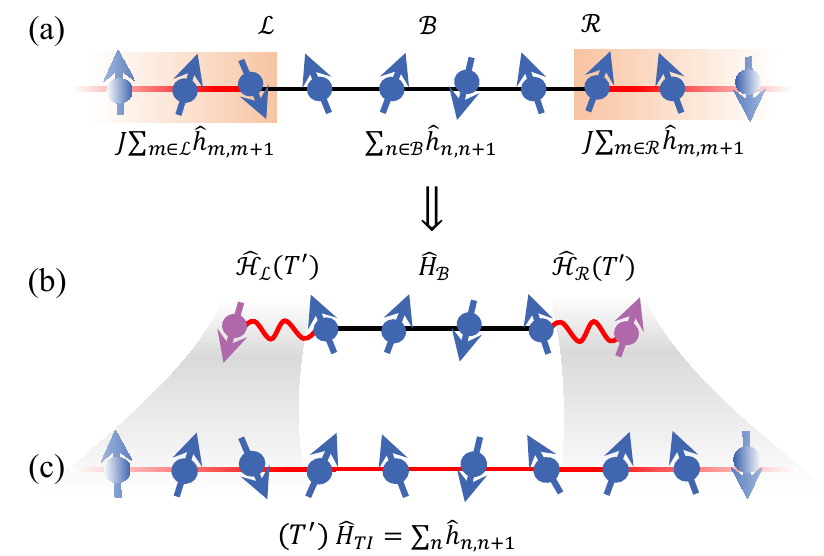}
	\caption{(Color online) Our goal is to construct a finite-size model whose bulk physics mimics that of the inhomogeneous system [Eq. (\ref{eq-Hinf})] shown in (a). Specifically, the interaction strengths in the bulk are taken as one and those outside the bulk as $J$. The idea is to variationally construct the boundary interaction $\hat{H}_{\mathcal{L}}(T')$ and $\hat{H}_{\mathcal{R}}(T')$ as shown in (b), where $T' = T/J$ is the effective \B{boundary} temperature. The boundary interactions are known as the entanglement-bath Hamiltonian that is optimized from the translational-invariant system $\hat{H}_{\text{TI}} = \sum_{n}\hat{h}_{n, n+1}$ \B{at finite temperature $T'$} as shown in (c).}
	\label{Fig1}
\end{figure}

\B{Let us start from} the quantum spin chain \B{that inspires this work, where the} interaction strengths are block-wisely inhomogeneous. The coupling strengths inside a finite-size block (namely the bulk denoted by $\mathcal{B}$) of \B{this} inhomogeneous system are set to be one as the energy scale, and the strengths in the rest of the system (i.e., the left and right environments denoted by $\mathcal{L}$ and $\mathcal{R}$, respectively) as $J$. The total Hamiltonian reads
\begin{equation}\label{eq-Hinf}
	\hat{H}_{\text{inho}}(J) = \sum_{n \in \mathcal{B}} \hat{h}_{n, n+1} + J \sum_{m \in \mathcal{L} \cup \mathcal{R}} \hat{h}_{m, m+1},
\end{equation}
with $\hat{h}_{n, n+1}$ the local two-body interaction, as illustrated in Fig. \ref{Fig1}(a). Note we set the interaction strengths between the bulk and environments also to be one, and suppose these terms are included in the bulk Hamiltonian $\sum_{n \in \mathcal{B}} \hat{h}_{n, n+1}$.

By taking $J$ to be different values, the bulk and environments could be in different phases, which will induce competition between the bulk and environments. \R{Such competition may cause novel physics. How to enhance or control this competition in a numerically- and experimentally-friendly way is an interesting and important issue. In this work, we introduce a few-body quantum entanglement simulator (QES) with only nearest-neighbor interactions, in which the infinite-size environments are renormalized into the two sites at the boundaries (see Fig. 1(c) to Fig. 1(b)). Thus, we expect QES to possess a strong competition between bulk and boundaries and induce exotic phenomena.} 

\R{Below, we describe the competition from the thermal perspective.} In inhomogeneous spin chain, $J$ can effectively tune the temperature difference between \B{the} bulk and environments. \B{Specifically}, we consider the finite-temperature density operator of $\hat{H}_{\text{inho}}$ satisfying 
\begin{eqnarray} \label{eq-rho}
	\hat{\rho}_{\text{inho}}(T;J) 
	&& = \exp{\left(-\frac{\hat{H}_{\text{inho}}(J)}{T}\right)} \B{/Z} \nonumber \\
	&& = \exp{\left(-\frac{\hat{H}_{\mathcal{L}}(J) + \hat{H}_{\mathcal{B}} + \hat{H}_{\mathcal{R}}(J)} {T}\right)} \B{/Z},
\end{eqnarray}
with the Hamiltonians $\hat{H}_{\mathcal{L}(\mathcal{R})}(J) = J \sum_{m \in \mathcal{L}(\mathcal{R})} \hat{h}_{m, m+1}$, $\hat{H}_{\mathcal{B}} = \sum_{n \in \mathcal{B}} \hat{h}_{n, n+1}$, and \B{the partition function $Z = \text{Tr} \left[e^{-\hat{H}_{\text{inho}}(J) / T} \right]$}. We take the Boltzmann constant $k_B=1$ for convenience. 

Defining an effective temperature
\begin{eqnarray} \label{eq-Tp}
	T' = \frac{T}{J},
\end{eqnarray}
we have
\begin{eqnarray} \label{eq-rho1}
	\ln \hat{\rho}_{\text{inho}}(T;T') = && -\frac{\sum_{m \in \mathcal{L}} \hat{h}_{m, m+1}}{T'} - \frac{\sum_{n \in \mathcal{B}} \hat{h}_{n, n+1}}{T} \nonumber \\&& - \frac{\sum_{m \in \mathcal{R}} \hat{h}_{m, m+1}}{T'} \B{- \ln Z}.
\end{eqnarray}
In other words, $J$ analogously adjusts the temperature of the environments, \B{and we expect that} the left and right environments act like the heat bath. The difference is that we still use the canonical distribution to describe the thermodynamics, which means all sites are physically in the same temperature \B{$T$}. We dub $T$ and $T'$ in the Hamiltonian of Eq. (\ref{eq-rho1}) as the \textit{physical} and \B{\textit{boundary}} temperature, respectively.

For $J=1$, the spin chain is translational invariant (TI), whose Hamiltonian reads
\begin{eqnarray} \label{eq-HTI}
\hat{H}_{\text{TI}} = \sum_{n}\hat{h}_{n, n+1}.
\end{eqnarray}
The thermodynamics of the infinite-size TI chain can be efficiently simulated by the existing tensor network (TN) methods~\cite{VMC08MPSPEPSRev, CV09TNSRev, 2017JPhA...50v3001B, 2020LNP...964.....R, RevModPhys.93.045003} such as the transfer-matrix renormalization group~\cite{PhysRevB.56.5061}, linearized tensor renormalization group~\cite{PhysRevLett.106.127202}, and TN tailoring~\cite{PhysRevB.105.155155}. For $J\neq 1$, the translational invariance is broken. The existing approaches become unstable when the system size is infinite. We here propose to \B{qualitatively mimic} the thermodynamics of the bulk $\mathcal{B}$ by constructing a finite-size model as illustrated in Fig. \ref{Fig1} (b). The key idea \B{is to put the EBHs at the boundaries. The EBHs are optimized with the infinite-size TI model at the temperature $T'$, which is expected to mimic the entanglement between the bulk and the environments in the thermal state of $\hat{H}_{\text{inho}}$.}

\B{To better explain our idea with the EBHs, let us consider a special case with $T'=T$. The constructed finite-size model becomes the quantum entanglement simulator (QES) whose bulk reduced density matrix (RDM) optimally mimics that of the infinite TI model~\cite{PhysRevE.93.053310, PhysRevB.96.155120} [Fig. \ref{Fig1} (c)]. The Hamiltonian of the QES consists of three parts as}
\begin{eqnarray} \label{eq-QES}
	\hat{H}_{\text{QES}}(T') =  \hat{\mathcal{H}}_{\mathcal{L}}(T') + \hat{H}_{\mathcal{B}} + \hat{\mathcal{H}}_{\mathcal{R}}(T').
\end{eqnarray}
The EBH $\hat{\mathcal{H}}_{\mathcal{L}(\mathcal{R})}(T')$ gives the interaction between the two spins on the left (right) boundary. It is obtained variationally by minimizing the distance between the entanglement Hamiltonians (EH's) of the QES and that of the TI model as
\begin{eqnarray} \label{eq-mini}
	\min_{\hat{H}_\mathcal{L}(T'), \hat{H}_\mathcal{R}(T')} \left|\hat{H}^{\text{E}}_{\text{TI}}(T') - \hat{H}^{\text{E}}_{\text{QES}}(T') \right|.
\end{eqnarray}
Note the EH of the bulk is defined as 
\begin{eqnarray} \label{eq-HE}
	\hat{H}^{\text{E}}_{\text{TI}}(T') = - \ln \text{Tr}_{/\mathcal{B}}  \left[ \exp\left(-\frac{\hat{H}_{\text{TI}}}{T'}\right) \B{/Z} \right],
\end{eqnarray}
with $\text{Tr}_{/\mathcal{B}}$ the trace of all degrees of freedom except for those of the bulk, and \B{ here the partition function $Z = \text{Tr} \left[e^{-\hat{H}_{\text{TI}} / T'} \right]$}. $\hat{H}^{\text{E}}_{\text{QES}}(T')$ is defined similarly by tracing over the boundary degrees of freedom. The sites on the very left and right sides of the QES are called the entanglement bath site (or bath in short), whose dimension $\chi$ is flexible. In the $\chi \to \infty$ limit, the distance given in Eq.~(\ref{eq-mini}) should approach zero. \R{Note that the numerical error given by Eq.~(\ref{eq-mini}) can be estimated by comparing the physical observables of QES and infinite-size system within the bulk. 
In this way, the excellent accuracy of QES for simulating infinite-size systems has been demonstrated in the previous works~\cite{PhysRevE.93.053310, PhysRevB.105.155155}.}

For an arbitrary $T'$, the EH of the bulk for the constructed model (we also call it QES below) at the physical temperature $T$ satisfies
\begin{eqnarray} \label{eq-HE1}
	\hat{H}^{\text{E}}_{\text{QES}}(T; T') = - \ln \text{Tr}_{/\mathcal{B}} \left[ \exp \left(-\frac{\hat{H}_{\text{QES}}(T')}{T}\right) \B{/Z} \right].
\end{eqnarray}
In an ideal case where the influence from the bulk to the infinite-size environments is ignorable even for $T\neq T'$, the minimization problem $\min_{\hat{H}_{\mathcal{L}(\mathcal{R})}(T')} \left|\hat{H}^{\text{E}}_{\text{inho}}(T; J) - \hat{H}^{\text{E}}_{\text{QES}}(T;T') \right|$ would have the same solution as Eq.~(\ref{eq-mini}). The QES constructed as Eq.~(\ref{eq-QES}) can accurately capture the influences from the environments to the bulk, and well mimic the bulk properties of $\hat{H}_{\text{inho}}$.

In practice, the influence from bulk to environment might not be ignorable, particularly when the bulk and the environments possess different physics. For instance, the bulk might possess the antiferromagnetic properties with a small $T$, and the environments might possess the paramagnetic physics with a large $T'$. Such cases with strong competition between the bulk and boundaries are exactly what we are interested in. Later we will show that the QES can still qualitatively mimic the bulk properties of $\hat{H}_{\text{inho}}$, i.e., the suppression of entropy, possibly due to the infinity of the environmental sizes. But we shall note that in spite of $\hat{H}_{\text{inho}}$, the QES itself and all its exotic properties, which are our main goals, are physically realizable when knowing all the terms in the EBHs (see, for example, Fig.~\ref{Fig3}). Thus, their significance does not have to rely on the accuracy of mimicking $\hat{H}_{\text{inho}}$. \R{As the EBH provides a proper approximation of the infinite-size environment, our results support the presence of the properties of $\hat{H}_{\text{QES}}$ (including the singularity discussed in the next section) to also appear in the infinite-size system $\hat{H}_{\text{inho}}$. But surely the properties of $\hat{H}_{\text{inho}}$ should be eventually verified by experiments. Anyway, we can experimentally realize $\hat{H}_{\text{QES}}$ directly to obtain all its exotic properties.}=

When calculating the EBHs, we take the coupling strength in the infinite-size TI chain as one [see Eq. (\ref{eq-HTI})]. Therefore, the $T'$ in Eqs. (\ref{eq-QES}) and (\ref{eq-mini}) equals to the boundary temperature $T'$ of the inhomogeneous system appearing in Eqs. (\ref{eq-Tp}) and (\ref{eq-rho1}). Then the QES could be described by the canonical distribution with a uniform physical temperature $T$ as Eq. (\ref{eq-rho}). The density operator of the QES at the physical temperature $T$ satisfies the canonical distribution as
\begin{eqnarray} \label{eq-rhoQES}
	\hat{\rho}_{\text{QES}}(T; T') = \exp \left( - \frac{\hat{H}_{\text{QES}}(T')}{T} \right) \B{/ Z}.
\end{eqnarray}
By tracing over all the degrees of freedom except for the bulk, the bulk properties are given by the RDM
\begin{eqnarray} \label{eq-RMD}
	\hat{\rho}^{\mathcal{B}}_{\text{QES}}(T; T') = \text{Tr}_{/\mathcal{B}} \hat{\rho}_{\text{QES}}(T; T').
\end{eqnarray}
The thermodynamic quantities, such as the von Neumann entropy per site we are interested in here, can be obtained as
\begin{eqnarray} \label{eq-ent}
	S_{n} = -\text{Tr}\hat{\rho}_{n}(T; T') \ln \hat{\rho}_{n}(T; T'), 
\end{eqnarray}
where $\hat{\rho}_{n}(T; T') = \text{Tr}_{/n}\hat{\rho}^{\mathcal{B}}_{\text{QES}}(T; T')$ is the $n$-site RDM, with $\text{Tr}_{/n}$ the trace of all degrees of freedom except for the $n$-th site.

\begin{figure}[tbp]
	\centering
	\includegraphics[angle=0,width=1\linewidth]{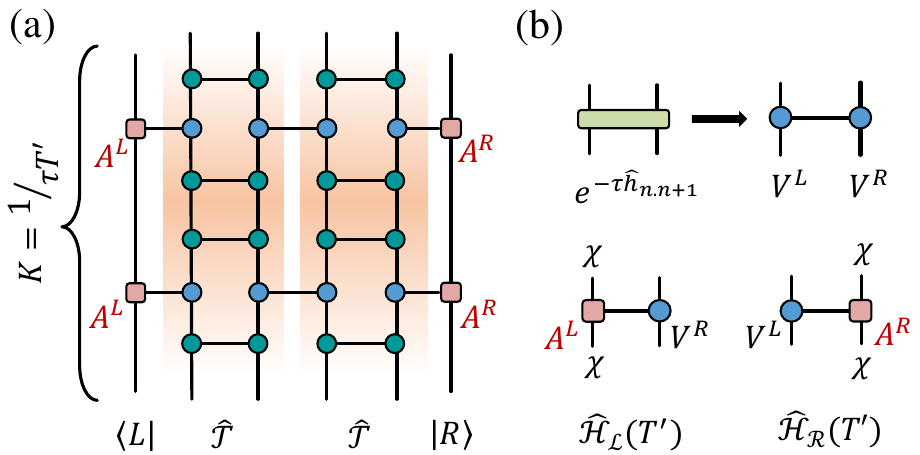}
	\caption{(Color online) (a) The thermal tensor network (TN) represents the density matrix of the TI chain at the temperature $T'$. The boundary MPS $|L\rangle$ and $|R\rangle$ can be calculated by the TN Tailoring approach. (b) The process of obtaining the EBHs. \B{$A^L$ or $A^R$ is inequivalent tensor, which forms the boundary MPS $|L\rangle$ or $|R\rangle$, with virtual bond dimension $\chi$. }}
	\label{Fig2}
\end{figure}

To obtain the EBHs at $T=0$ (the ground state), i.e., $\hat{\mathcal{H}}_{\mathcal{L}(\mathcal{R})}(T'=0)$, one can use the existing methods such as infinite density matrix renormalization group and tensor ring encoding~\cite{PhysRevE.93.053310}. Non-trivial phenomena of the ground state are exhibited in the bulk by tuning the EBHs~\cite{PhysRevB.98.085111}. For the EBHs at finite temperatures, we employ the TN Tailoring approach~\cite{PhysRevB.105.155155}, which gives the stay-of-the-art accuracy in simulating the thermodynamics of one-dimensional many-body systems. Specifically, we construct the TN that represents the density operator $\exp\left(-\frac{\hat{H}_{\text{TI}}}{T'}\right)$. Then we calculate the boundary matrix product states (MPS) $|L\rangle$ and $|R\rangle$, which are the left and right dominant eigenstates of the transfer matrix $\hat{\mathcal{T}}$ of the TN, respectively [Fig. \ref{Fig2}(a)]. We have $\lambda^* |L\rangle = \hat{\mathcal{T}}^{\dagger} |L\rangle$ and $\lambda |R\rangle = \hat{\mathcal{T}} |R\rangle$ with $\lambda$ the eigenvalue. In other words, the boundary MPS's are optimized by maximizing the free energy per site $f$ as
\begin{eqnarray} \label{free energy}
	f = \max_{|L\rangle,|R\rangle} \left(-T'  \ln \lambda \right).
\end{eqnarray}

The boundary MPS's are uniform (namely formed by the copies of one inequivalent tensor, denoted by $A^L$ and $A^R$, respectively), and possess the periodic boundary condition along the temporal direction. The virtual bond dimension $\chi$ of $A^L$ and $A^R$, which determines the dimension of the entanglement bath sites mentioned above, is relating to the performance of the EBHs for mimicking the infinite-size TI model at the target temperature. The larger the virtual bond dimension is, the higher the accuracy of the QES will be [Eq.~(\ref{eq-mini})]~\cite{PhysRevB.105.155155}. The length of the boundary MPSs satisfies $K=\frac{1}{\tau T'}$ with $\tau$ a small positive number known as the Trotter slice. The EBHs are obtained by $A^{L(R)}$ and $V^{L(R)}$ as illustrated in Fig. \ref{Fig2}(b), where $V^{L(R)}$ are obtained by decomposing the local imaginary-time evolution operator $\exp \left(-\tau \hat{h}_{n, n+1}\right)$. \R{It should be noted that different models correspond to different boundary MPS obtained through the TN methods. Therefore, EBH should be model-dependent.}

\section{Exotic bulk entropy induced by entanglement-bath Hamiltonians}

\subsection{Interactions in the entanglement-bath Hamiltonians}
\begin{figure}[tbp]
	\centering
	\includegraphics[angle=0,width=0.8\linewidth]{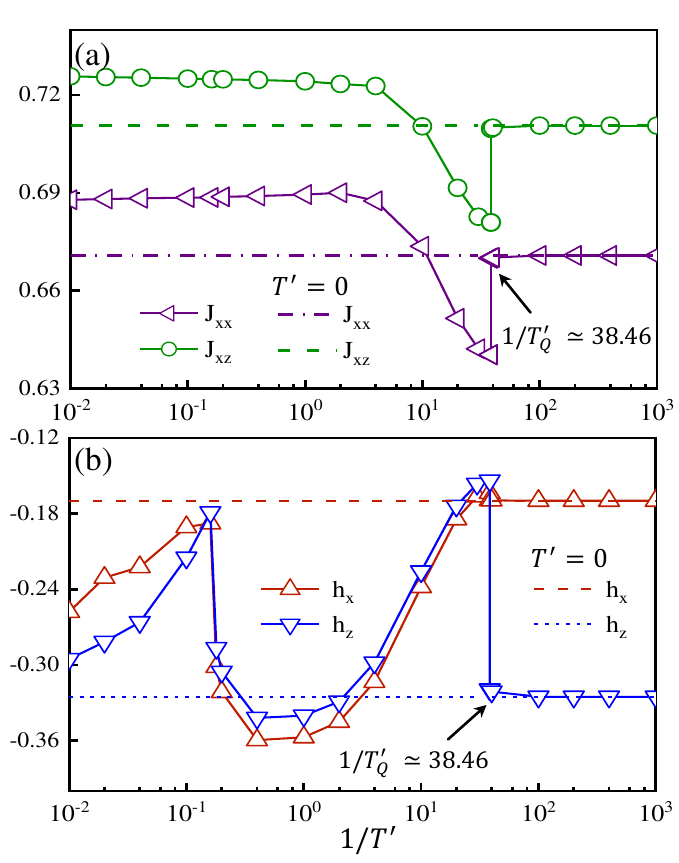}
	\caption{(Color online) (a) The coupling strength \B{$J^{\alpha x}_{\mathcal{L}(\mathcal{R})}(T')$} and (b) the magnetic fields \B{$h^{\alpha}_{\mathcal{L}(\mathcal{R})}(T')$} of the EBHs \B{[see Eq.~(\ref{eq-HLR})]} versus the inverse boundary temperatures $1/T'$. The dash lines give the couplings at the zero boundary temperature $T'=0$~\cite{PhysRevB.98.085111}. The discontinuous point indicates the \B{boundary} quench point $T'_Q$.}
	\label{Fig3}
\end{figure}

We take the quantum Ising chain as an example, where the local Hamiltonian satisfies
\begin{equation} \label{eq-Ising}
	\hat{h}_{n, n+1} = \hat{S}^z_n \hat{S}^z_{n+1} - \frac{h}{2} (\hat{S}^x_n + \hat{S}^x_{n+1}),
\end{equation}
with $h$ the transverse magnetic field. \R{For $\chi=2$, the EBHs on the two boundaries can be expanded by the interactions of spin-$1/2$'s as}
\begin{eqnarray}\label{eq-HLR}
&&\hat{\mathcal{H}}_\mathcal{L}(T') = \sum_{\alpha=x,z}[J^{\alpha x}_{\mathcal{L}}(T') \hat{S}_{1}^{\alpha} \hat{S}_{2}^{x} - \frac{1}{2} h^{\alpha}_{\mathcal{L}}(T') \hat{S}_{1}^{\alpha}], \nonumber \\
&&\hat{\mathcal{H}}_{\mathcal{R}}(T') = \sum_{\alpha=x,z}[J^{x \alpha}_{\mathcal{R}}(T') \hat{S}_{N+1}^{x} \hat{S}_{N+2}^{\alpha} - \frac{1}{2} h_{\mathcal{R}}^{\alpha}(T') \hat{S}_{N+2}^{\alpha}]. \nonumber\\
\end{eqnarray}
The coupling terms are consistent with those \R{for the ground state reported in Ref.~[\onlinecite{PhysRevB.98.085111}] (see the horizontal dashed lines in Fig.~\ref{Fig3}), except that the coefficients here rely on $T'$ (see the solid lines with symbols). We shall recall that the EBHs located on the left and right boundaries are numerically obtained using the TN tailoring method \cite{PhysRevB.105.155155} to simulate the infinite-size TI chain at the temperature $T'$. Thus, in this stage of calculating the EBHs, we take $T=T'$.} 

Besides the coupling constants $\hat{S}^z \hat{S}^z$ and the transverse field $\hat{S}^x$ that originally exist in the quantum Ising model, the $\hat{S}^x \hat{S}^z$ coupling and a vertical field \B{$h_{\mathcal{L}(\mathcal{R})}^z$} emerge in the EBHs. Due to the symmetries of the system, we also obtain
\begin{eqnarray}\label{eq-J}
	&J_{\mathcal{L}}^{xz} = -J_{\mathcal{R}}^{zx} \doteq J_{xz},\ \  h_{\mathcal{L}}^{x} = h_{\mathcal{R}}^{x} \doteq h_{x}, \\
	&J_{\mathcal{L}}^{xx} = -J_{\mathcal{R}}^{xx} \doteq J_{xx},\ \  h_{\mathcal{L}}^{z} = h_{\mathcal{R}}^{z} \doteq h_{z}. 
\end{eqnarray}
The coupling strengths have odd parity while magnetic fields exhibit even parity when changing from the left to the right environment. Same as the ground-state cases, this is because the couplings are antiferromagnetic, and the magnetic field is uniform.

The difference from the ground-state case is that the strengths of the couplings here vary with $T'$. See Fig. \ref{Fig3} for the coupling strengths and magnetic fields in the EBHs at $h=0.5$ [Eq. (\ref{eq-Ising})] with different $T'$. Note that the spins-$1/2$'s on the boundaries of the QES \B{corresponds $\chi=2$ on the boundary MPS's}, and in principle, it can be replaced by the high-dimensional spins~\cite{PhysRevE.93.053310, PhysRevB.96.155120} \R{with $\chi >2$.}

In comparison with the realization of the bulk thermodynamics of the inhomogeneous system $\hat{H}_{\text{inho}}(J)$ [Eq. (\ref{eq-Hinf})], an advantage of employing the QES scheme is that it concerns the bounded strengths of the couplings. For $\hat{H}_{\text{inho}}(J)$, $J$ increases linearly as $T'$ lowers [Eq. (\ref{eq-Tp})]. Therefore, \B{a low $T'$} requires a large coupling strength $J$ which is challenging for the experiment realization. With the QES, the coupling strengths in the EBHs are bounded. Down to the low $T'$, the strengths of all terms in the EBHs converge to some finite values with the order of magnitude $O(10^{-1})$. 

Moreover, a discontinuous point is observed at \R{$1/T' \simeq 38.46$}, which we dub as the \textit{boundary quench point} (BQP) (denoted by $T'_Q$). \R{It can be observed straightforwardly in Fig.~\ref{Fig3} that for $ T' < T'_Q $, the coupling strengths of EBHs become approximately identical to those obtained in the ground state (see the dash lines). It means that for $ T' < T'_Q $, the bath is dominated by the ground-state physics, and the effects of the bath to the bulk become the same as those of the bath obtained for the ground-state simulation. Such a ``transition'' indicated by the BQP is driven by the $T'$ of the environments, differing from the thermodynamic phase transitions that are driven by thermal fluctuations of the whole system.} We will come back to the BQP below with more properties from the perspective of bulk entropy.

\subsection{Bulk entropy and boundary quench point}
\begin{figure}[tbp]
	\centering
	\includegraphics[angle=0,width=0.85\linewidth]{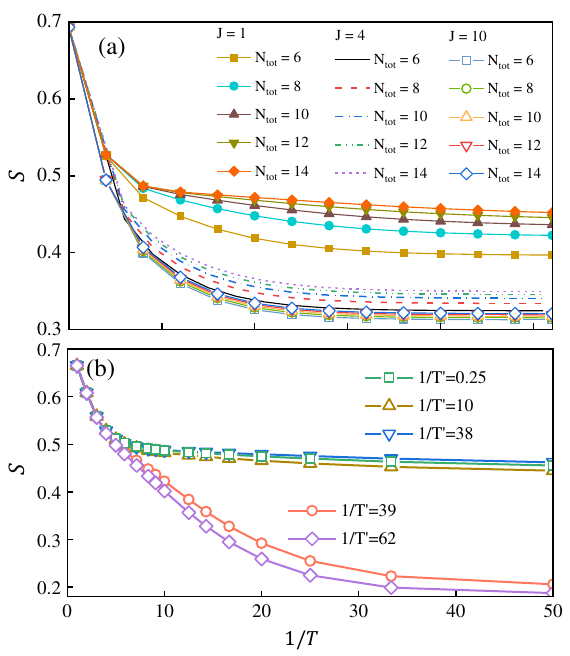}
	\caption{(Color online) \B{(a) The average bulk entropy $S$ by ED versus the inverse temperatures $1/T$ for the inhomogeneous system with different system size $N_{\text{tot}}$ and  different coupling strength $J$ [Eq.~(\ref{eq-Hinf})]. (b) The $S$ of the QES with different $T'$ [Eq. (\ref{eq-rhoQES}) or (\ref{eq-HLR})]. By increasing the boundary temperature from $1/T'=38$ to $39$ that crosses the BQP, the convergent value of the entropy is altered from $S\simeq0.44$ to $0.18$.}}
	\label{Fig4}
\end{figure}
Fig. \ref{Fig4} compares the average bulk entropy $S = \sum_{n=1}^{N}S_n / {N}$ of the inhomogeneous system $\hat{H}_{\text{inho}}(J)$ and that of the QES $\hat{H}_{\text{QES}}(T')$. The $T'$ in the EBHs of the QES and the $J$ in the environments of the inhomogeneous system satisfy Eq. (\ref{eq-Tp}). 

We use the exact diagonalization (ED) to obtain the entropy of the inhomogeneous system by varying $J$ and taking the total size $N_{\text{tot}}$ from $6$ to $14$ as shown in Fig. \ref{Fig4}(a). The bulk size is fixed to be $N=4$. The average bulk entropy $S$ decreases as the physical temperature $T$ lowers, as expected. For all sizes, we observe that $S$ is suppressed by increasing $J$ (meaning lowering the boundary temperature). This is analogous to the phenomenon when lowering the boundary temperature in an open system. No essential change of such a phenomenon can be seen from our results if we further increase the size of the environments.


For the $\hat{H}_{\text{QES}}(T')$ [Fig. \ref{Fig4}(b)], similar suppression of $S$ by $T'$ is observed. This supports our conjecture that adjusting the parameters in the EBHs by following the results given in Fig.~\ref{Fig3} will qualitatively realize the bulk physics induced by tuning $J$ in the inhomogeneous system. For a relatively small $J$ (say $J=1$), the entropy of $\hat{H}_{\text{inho}}$ converges to $S \simeq 0.46$ as $N_{\text{tot}}$ increases for $T \to 0$. This coincides with the entropy of the QES with a relatively small $1/T'$, where we have $S \simeq 0.46$ as for $T \to 0$. For a large $J$ in $\hat{H}_{\text{inho}}$ where the spins in the environments intend to be ordered, the entropy is suppressed to be $S \simeq 0.31$ for $J=10$ and $T \to 0$. For the QES, the suppression is more drastic by lowering $T'$. We have $S \simeq 0.19$ for $1/T'=62$ and $T \to 0$. More interesting, a significant change on the convergent value of $S$ occurs between $1/T'=38$ and $39$, in which the BQP locates. There is no essential change of such a phenomenon when bulk size is increasing.

\begin{figure}[tbp]
	\centering
	\includegraphics[angle=0,width=0.85\linewidth]{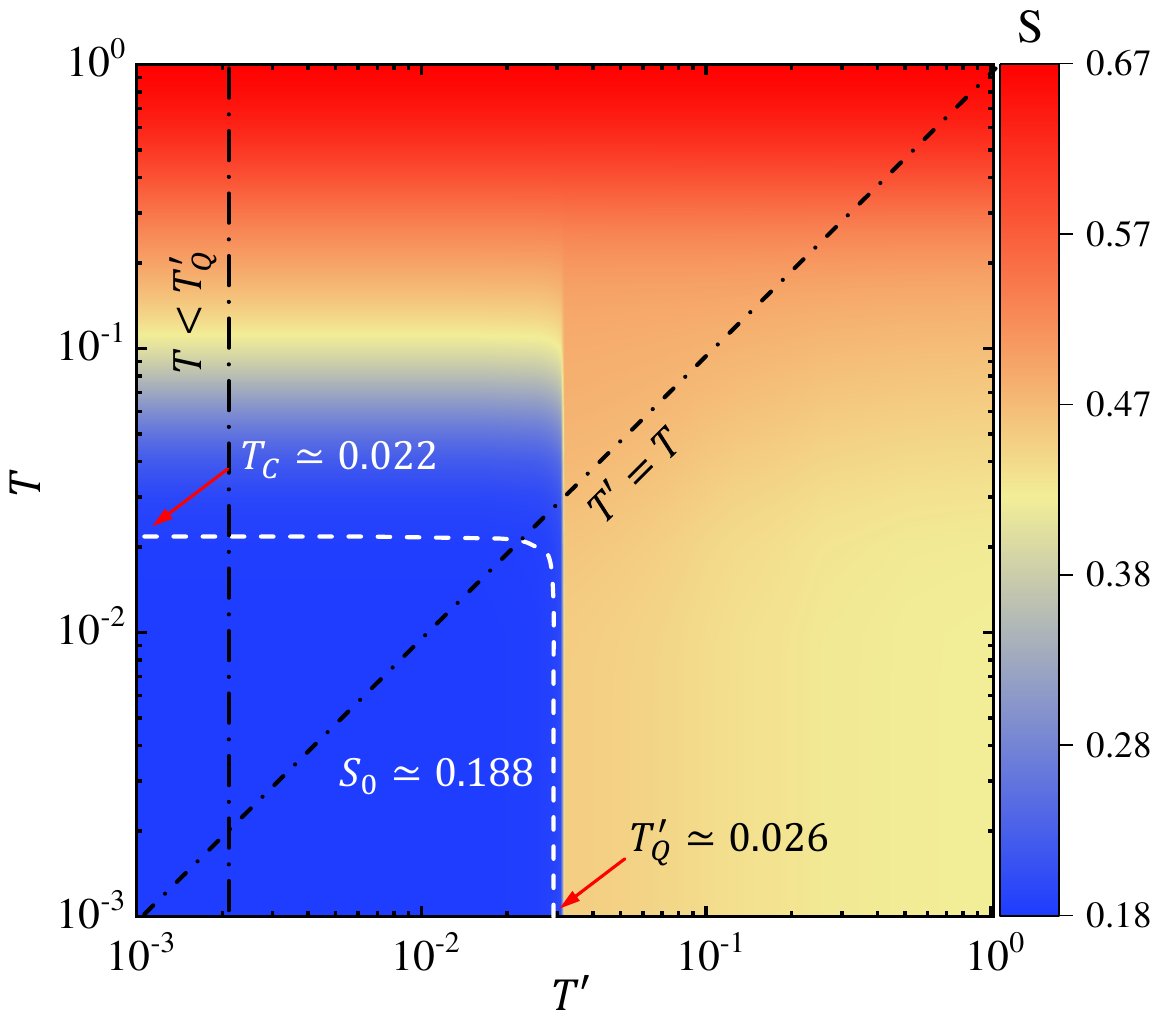}
	\caption{(Color online) The average bulk entropy $S$ of the QES for the quantum Ising model at different physical temperatures $T$ and boundary temperatures $T'$. We take the transverse field at the critical point $h_x=0.5$. $S$ is suppressed to $S=0.188$ for about $T<0.022$ and $T'<0.026$.}
	\label{Fig5}
\end{figure}

The above results suggest that the difference between the inhomogeneous system $\hat{H}_{\text{inho}}(J)$ and QES $\hat{H}_{\text{QES}}(T')$ is the smoothness against $T'$ (or $J$). Fig.~\ref{Fig5} demonstrates the $S$ in the QES versus the physical and boundary temperatures ($T$ and $T'$, respectively). The bulk size is taken as $N=8$. We shall stress that even for $T \neq T'$, the temperature defining the canonical distribution is the physical temperature $T$ [Eq. (\ref{eq-rhoQES})]. The boundary temperature $T'$ is a hyper-parameter of the couplings in the EBHs (see Eq. (\ref{eq-HLR}) and Fig. \ref{Fig3}). The drop of $S$ is much more drastic for $T' < T'_Q$ compared with those for $T' > T'_Q$, with $T'_Q$ the BQP. The reason is that the EBHs for $T' < T'_Q$ jumps to those of the ground state (analog to the zero-temperature bath), while the EBHs for $T' > T'_Q$ possess obvious $T'$-dependence (see Fig.~\ref{Fig3}). 

When $T'$ and $T$ are both low, the bulk entropy $S$ is suppressed to a low value with $S \simeq S_0\simeq0.188$, as demonstrated by the blue area \B{in the bottom-left corner}. Define the \textit{thermal cross-over point} (TCP) $T_{C}$ as the temperature, where we have $S\simeq S_{0}$ for $T<T_{C}$. Our result shows $T'_Q \simeq T_{C}+ O(10^{-3})$, which indicates the underlying equivalence of the scalings between these two temperatures. This can be explained by considering $T=T'$. In this case, the QES optimally mimics the infinite-size TI model $\hat{H}_{\text{TI}}$, where the cross-over temperature is accurately predicted by the drop of $S$. Therefore, the BQP and TCP should be unified to the same value. 

\begin{figure}[tbp]
	\centering
	\includegraphics[angle=0,width=0.95\linewidth]{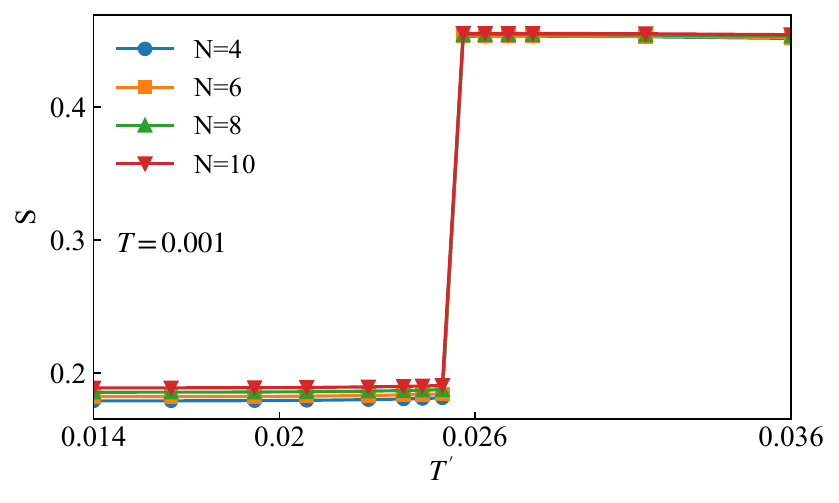}
	\caption{\R{(Color online) The average bulk entropy $S$ versus the boundary temperature $T'$ for different bulk sizes ($N=4, \ldots, 10$). We here fix $T=0.001$.}}
	\label{FigN}
\end{figure}

\R{Fig.~\ref{FigN} demonstrates the robustness of the BQP for different bulk sizes $N$. We take $T=0.001$, meaning the data corresponds to the horizontal path in Fig.~\ref{Fig5} at the bottom. By changing $N$ from $4$ to $10$, the singularity and the position ($T'_{\text{Q}}$) of the BQP stay almost unchanged.}

\begin{figure}[tbp]
	\centering
	\includegraphics[angle=0,width=0.95\linewidth]{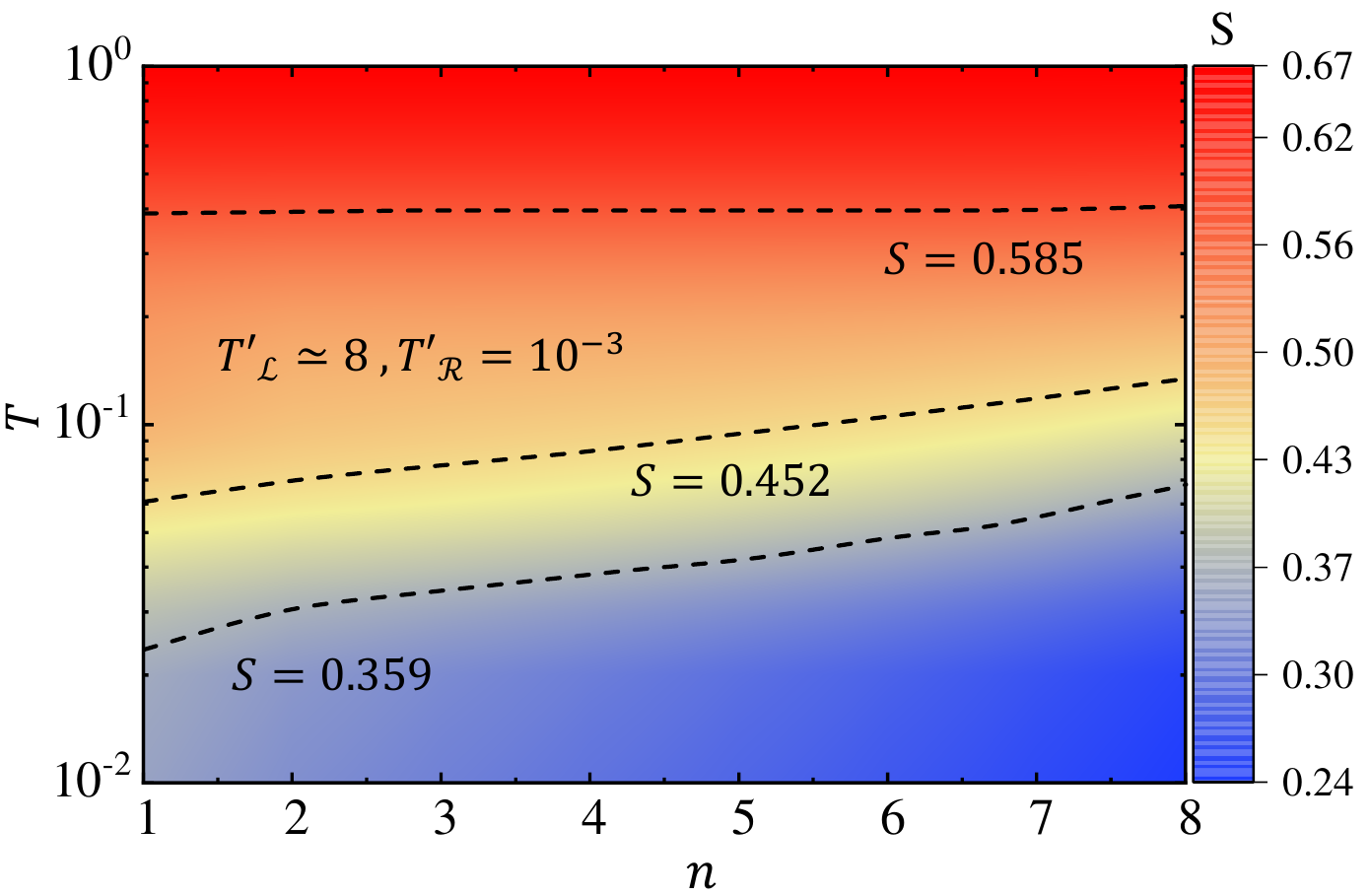}
	\caption{(Color online) The on-site entropy $S_n$ of the QES for the quantum Ising model on different site $n$ of the bulk at different physical temperatures $T$. Taking the left and right bath temperatures as $T'_{\mathcal{L}} \simeq 8$ and $T'_{\mathcal{R}} \simeq 10^{-3}$, respectively, a slope of the spatial distribution of the fluctuations is observed.}
	\label{Fig6}
\end{figure}

\R{The boundary temperatures for the left and right EBHs can be taken differently}. Fig. \ref{Fig6} shows the on-site entropy $S_n$ [Eq.~(\ref{eq-ent})] at different sites ($n$) with the left and right boundary temperatures as $T'_{\mathcal{L}} = 8$ and $T'_{\mathcal{R}} = 10^{-3}$, respectively. The physical temperature is still uniformly $T$ and the system is still described by the canonical distribution as Eq. (\ref{eq-rhoQES}). The spatial distribution of $S_n$ shows a non-zero gradient in the bulk of the QES. It indicates that the EBH with a low (high) $T'$ tends to drive the system into an ordered (disordered) state, similar to a heat bath with a low (high) temperature. \B{It is interesting to note such a phenomenon usually appears in a grand canonical ensemble~\cite{RevModPhys.61.981}, while the above results are obtained from a canonical ensemble}.

\subsection{\B{Boundary quench point and symmetries of Hamiltonian}}

\begin{figure}[tbp]
	\centering
	\includegraphics[angle=0,width=0.85\linewidth]{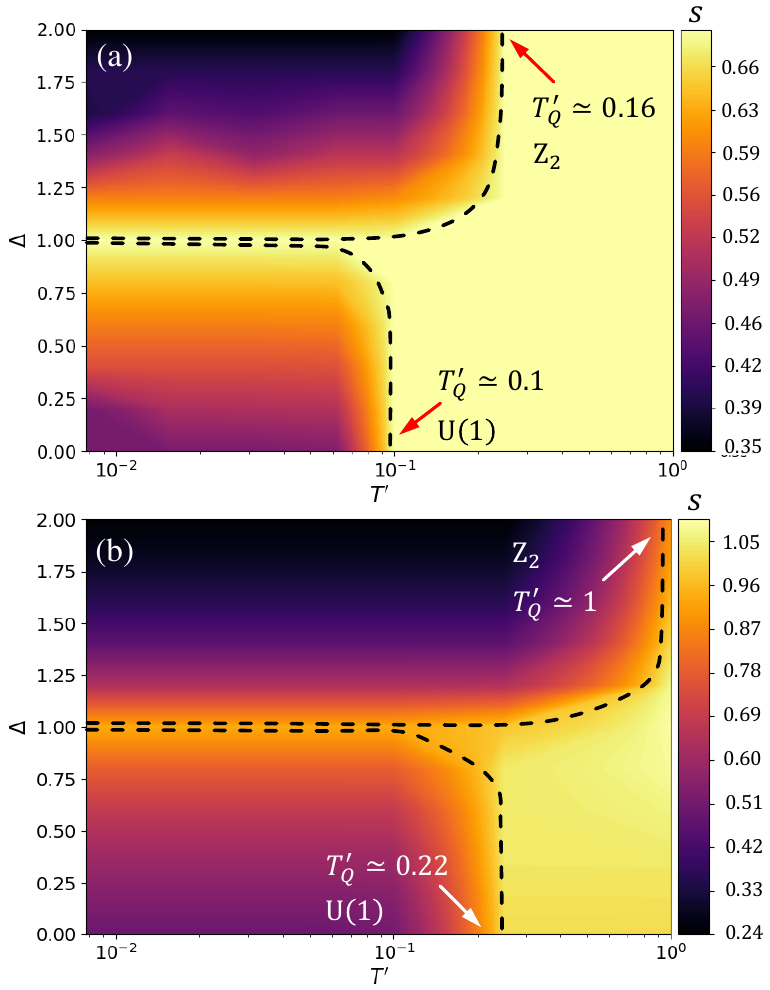}
	\caption{(Color online) \B{The average bulk entropy $S$ for the XXZ chains varies with different anisotropy parameters $\Delta$ and boundary temperature $T'$. (a) The spin-1/2's case, BQP exists at $\Delta\ne1$ (SU(2) symmetry broken) with $T'_Q\simeq0.1$ at $0 \leq \Delta < 1$ (XY phase) and $T'_Q\simeq0.16$ at $\Delta > 1$ (Ising antiferromagnetic phase).  (b) The spin-1's case, BQP exists at $\Delta\ne1$ ($SU(2)$ symmetry broken) with $T'_Q\simeq0.22$ at $0 \leq \Delta < 1$ and $T'_Q \simeq 1$ at $\Delta > 1$.}}
	\label{Fig7}
\end{figure}
\B{Taking the spin-1/2 and spin-1 XXZ chains as examples, we investigate the relations between the BQPs and the symmetries of Hamiltonian. The Hamiltonian of the XXZ model is written as}
\begin{align}\label{XXZ}
	\hat{H} = \sum_{n}(\hat{S}^{x}_n\hat{S}^{x}_{n+1} + \hat{S}^{y}_n\hat{S}^{y}_{n+1}+\Delta\hat{S}^{z}_n\hat{S}^{z}_{n+1}), 
\end{align}
\B{where $\Delta$ is the spin anisotropy parameter. For $\Delta=1$, the XXZ chain becomes the Heisenberg chain satisfying the SU(2) symmetry. The average bulk entropy $S$ versus the boundary temperature $T'$ and anisotropy parameter $\Delta$ is shown in Fig.~\ref{Fig7} (a) for the spin-1/2 case and in Fig.~\ref{Fig7} (b) for the spin-1 case, where the physical temperature is fixed to be $0.01$ [Eq. (\ref{eq-QES})]. Different symmetries lead to different entropic behaviors. Specifically, for both the spin-1/2 and spin-1 cases with $\Delta=1$ where the SU(2) preserves, $S$ is almost independent of $T'$ and the BQP would not appear.}
	
\R{For $\Delta \ne 1$, the Hamiltonian with SU(2) symmetry is reduced to the XY-type ($\Delta < 1$) or Ising-type ($\Delta > 1$), which is dominated by U(1) or Z$_{2}$ symmetry, respectively.} The system undergoes a quantum phase transition between the XY phase and the Ising-type antiferromagnetic phase. The BQP emerges and its position remains approximately unchanged within each phase. We have $T'_{Q} \simeq 0.1$ in the XY phase and $T'_{Q} \simeq 0.16$ in the antiferromagnetic phase for the spin-1/2 chain, and $T'_{Q} \simeq 0.22$ in the XY phase and $T'_{Q} \simeq 1$ in the antiferromagnetic phase for the spin-1 chain. When tuning the boundary temperature to be above the BQP ($T'>T'_{Q}$), the system enters the high-temperature paramagnetic phase. These results uncover the underlying connections between the BQP and the symmetry breaking.

\subsection{\B{Boundary effect from finite bath dimension}}

\begin{figure}[tbp]
	\centering
	\includegraphics[angle=0,width=0.8\linewidth]{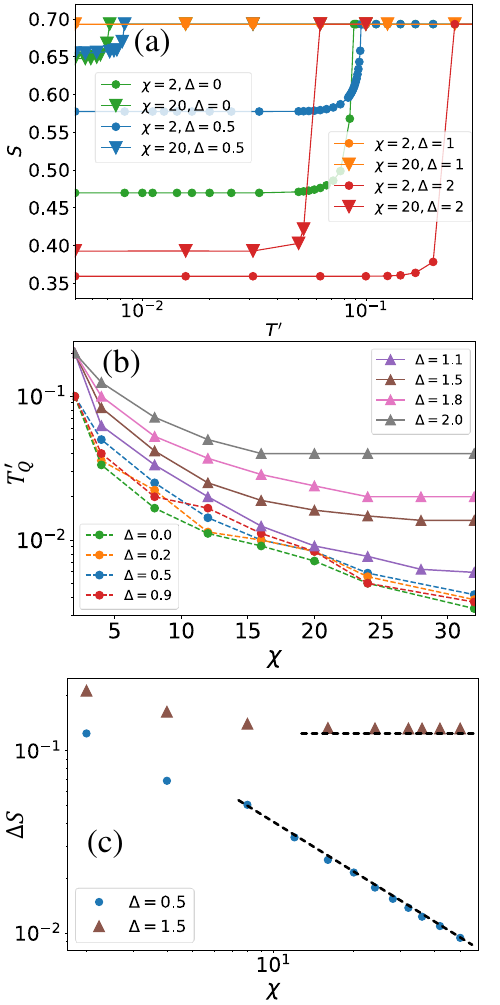}
	\caption{(Color online) (a) The average bulk entropy $S$ for spin-1/2 XXZ chain versus $T'$ with \R{different bath dimensions ($\chi=2$ and $20$) and anisotropy parameters ($\Delta=0, 0.5, 1, 2$).} (b) The BQP $T'_Q$ versus $\chi$ \R{for different $\Delta$}. (c) The jump of bulk entropy $\Delta S$ versus $\chi$ \R{at the BQP for $\Delta=0.5$ and $1.5$.}}
	\label{Fig8}
\end{figure}

For simulating the bulk properties of the infinite-size TI system (namely $T = T'$), the infinite-size environments are renormalized into the $\chi$-dimensional baths at the boundaries of the QES [see Fig. \ref{Fig1} (c) to (b)]. The baths can dramatically lower the finite-size effects, where the bath dimension $\chi$ determines the accuracy of the QES [Eq.~(\ref{eq-mini})] and gives the upper bound $\ln\chi$ of the entanglement entropy between the bulk and each bath in the QES. In the $\chi \to \infty$ limit, the distances between infinite-size environments and baths should approach to zero and the bulk properties of infinite chains can be given exactly by the QES. 

We show the influence of the finite bath dimension with fixed bulk size $N=4$. Fig. \ref{Fig8} (a) demonstrates the average bulk entropy $S$ versus the boundary temperature $T'$ with $\chi=2$ and $20$ with a low physical temperature $T = 0.01$. For both $\chi$'s, $S$ exhibits the discontinuity at the BQP for $\Delta =0$, $0.5$, and $2$. No BQP appears for $\Delta=1$. The position of the BQP moves to the lower $T'$ as $\chi$ increases. This is possibly because that enlarging the bath dimension could increase the entanglement and thus the competitions between the baths and bulk. Consequently, larger $\chi$ allows the baths to induce the BQP at a lower temperature. Meanwhile, larger $S$ is obtained for $T' < T'_{Q}$ for the same reason. 

\R{Fig.~\ref{Fig8} (b) shows the BQP $T'_{Q}$ versus $\chi$ with different anisotropy parameter $\Delta$. It is obvious that for $\Delta < 1$ (XY-type Hamiltonian), the BQP occurs at a lower temperature as $\chi$ increasing, while $T'_{Q}$ converges with $\chi$ for $\Delta > 1$ (Ising-type Hamiltonian). The converged $T'_{Q}$ decreases when approaching to the Heisenberg point $\Delta \to 1$ from the Ising side. Fig.~\ref{Fig8} (c) shows the height of jump of $S$ at the BQP $\Delta S = S_{{T'_{Q}}^{(+)}} - S_{{T'_{Q}}^{(-)}}$ with different values of $\chi$. Here, we take $\Delta=0.5$ and $1.5$ as two examples, where the Hamiltonian is XY-type and Ising-type, respectively. It demonstrates an algebraic scaling between $\Delta S$ and $\chi$ for $\Delta=0.5$, while $\Delta S$ converges at $0.13$ approximately for $\Delta=1.5$ as $\chi$ increases. The results above indicates that in the large $\chi$ limit, the BQP seems to disappear in the XY phase, whereas it is robust in the Ising-type antiferromagnetic phase.}

\AtEndEnvironment{thebibliography}{
	\bibitem{pytorch} The official website of PyTorch is at \url{https://pytorch.org/}
}

\section{EXPERIMENTAL IMPLEMENTATIONS}

Reducing $\hat{H}_{\text{inho}}(J)$ to the QES is advantageous to its experimental realization. The EBH contains just one-body and two-body terms, and their strengths are bounded even for the extremely low boundary temperatures \B{(or quantitatively large $J$)}. For the case of spin-$1/2$ models, one needs the controlled coupling between the boundary spins and auxiliary (non-necessarily $1/2$) spins. We foresee four possible platforms to realize these tasks. Each of these platforms contributes to the pillars of contemporary Quantum Technologies: quantum computing, quantum simulation, and quantum metrology~\cite{Acin18}. We list and comment on them below:
\begin{itemize}
	
	\item {\it Ultracold trapped ions.} This is probably the most promising platform, so we focus on it. These systems reduce to spin models, in which ions' internal states serve spin states, and couplings are mediated by phonons/lasers (cf.~[\onlinecite{DavidNatComm,DavidPRA}] and references therein; for an overview of the underlying theory and experiments see~\cite{Mintert,Porras,Schaetz08,Schatz16a,Schaetz16b,Jurcevic17,Li17,Monroe17,Zhang17,Bollinger17,Monroe21}). One could design additional traps at the edges of the system with the same or even different ions and couple them in the desired controlled way to the bulk. Such an approach would work even in two dimensions.
	
	\item {\it Trapped Rydberg atoms.} Similarly, one could use arrays of trapped Rydberg atoms that may serve as simulators of spin models with long-range couplings~\cite{Lukin,Broaways,Broaways1,Lukin21}. Again,  the idea is to design additional traps with auxiliary (generally different) Rydberg atoms, and design couplings of these atoms to the bulk by, e.g., following the data in Fig.~\ref{Fig2}.
	
	\item {\it Ultracold atoms in optical lattices.} Spin models can be realized with ultracold atoms in optical lattices employing for instance super-exchange interactions (cf.~[\onlinecite{LSA17}]). Using contemporary super-lattice/holographic mask techniques, one can design a lattice, in which atoms in other internal states are trapped, and are brought to interact with the bulk in the desired way.
	
	\item{\it Ultracold atoms in nano-structures.} Such systems realize spin models with controlled long-range interactions~\cite{Darrick1,Darrick2,Darrick3}. Again, an appropriate design of nano-structures allows one to add additional traps and atoms at the edges.
	
\end{itemize}

\section{Summary and Discussion}

In summary, we expose the exotic boundary-induced thermodynamic properties with the entanglement bath Hamiltonian (EBH). The EBH is variationally determined as the Hamiltonian that reproduces the bulk entanglement Hamiltonian of the infinite-size translational-invariant system at finite temperatures. We show that bulk entropy can be controlled by tuning the EBH. A discontinuous point BQP is found on the coefficients of the EBH and the bulk entropy. It indicates a low-entropy region where the thermal fluctuations are suppressed to be insignificant and the bulk properties are dominated by the ground-state physics. Taking spin-1/2 and spin-1 chains as examples, we show that the BQP \R{disappears at the Heisenberg point with full SU(2) symmetry, but emerges when reducing to the XY-type (dominated by U(1) symmetry) or Ising-type (dominated by Z$_{2}$ symmetry) Hamiltonian. Moreover, the impacts from the entanglement-bath dimension are investigated. The BQP approaches to the zero temperature in the large $\chi$ limit in the XY phase, while it is robustly at a finite temperature in the Ising-type antiferromagnetic phase.} The possible experimental realizations of the uncovered boundary physical phenomena are discussed.

Our results suggest that the boundary quench point (BQP) reveals a novel thermodynamic phenomenon that differs from the conventional phase transitions. The thermodynamic phase transitions are normally driven by thermal fluctuations at finite temperatures. For the quantum phase transitions, the fluctuations are from the competition between different phases. \R{The ``transition'' indicated by a BQP is also caused by the fluctuations in the competition of two phases. The essential difference is that such competition is between the bulk and environments (entanglement baths)}. Thus, a BQP could be regarded as the result of boundary-bulk competitions, which could play an important role in understanding the quantum many-body physics with non-trivial boundaries. With the thermodynamic TN methods such as projected entangled pair operator~\cite{RLXZS12ODTNS, CCD12FTPEPS, RXLS13NCD, CD15TPO, KWO17MPSopen, KREO18PEPOthermal}, we may explore the boundary-induced singularity in higher dimensions.

\begin{acknowledgments}
	The authors are grateful to Gang Su, Wei Li, Han Li, Kai Xu, Han-Jie Zhu, Bin-Bin Chen and Leticia Tarruell for stimulating discussions. This work is supported by National Nature Science Foundation of China (No. 12004266, No. 11834014 and Grant No. 12074027), Beijing Natural Science Foundation (Grant No. 1232025), and the key research project of Academy for Multidisciplinary Studies, Capital Normal University. 
	
	ML acknowledges the support from: ERC AdG NOQIA; MICIN/AEI (PGC2018-0910.13039/501100011033, CEX2019-000910-S/10.13039/501100011033, Plan National FIDEUA PID2019-106901GB-I00, FPI; 
	MICIIN with funding from European Union NextGenerationEU (PRTR-C17.I1): QUANTERA MAQS PCI2019-111828-2); 
	MCIN/AEI/10.13039/501100011033 and by the “European Union NextGeneration EU/PRTR"  QUANTERA DYNAMITE PCI2022-132919 (QuantERA II Programme co-funded by European Union’s Horizon 2020 programme under Grant Agreement No 101017733), Ministry of Economic Affairs and Digital Transformation of the Spanish Government through the QUANTUM ENIA project call – Quantum Spain project, and by the European Union through the Recovery, Transformation and Resilience Plan – NextGenerationEU within the framework of the Digital Spain 2026 Agenda; 
	Fundació Cellex; Fundació Mir-Puig; Generalitat de Catalunya (European Social Fund FEDER and CERCA program, AGAUR Grant No. 2021 SGR 01452, QuantumCAT \ U16-011424, co-funded by ERDF Operational Program of Catalonia 2014-2020); Barcelona Supercomputing Center MareNostrum (FI-2023-1-0013); EU Quantum Flagship (PASQuanS2.1, 101113690); EU Horizon 2020 FET-OPEN OPTOlogic (Grant No 899794); EU Horizon Europe Program (Grant Agreement 101080086 — NeQST), National Science Centre, Poland (Symfonia Grant No. 2016/20/W/ST4/00314); ICFO Internal ``QuantumGaudi'' project; ``La Caixa'' Junior Leaders fellowships ID100010434: LCF/BQ/PI19/11690013, LCF/BQ/PI20/11760031, LCF/BQ/PR20/11770012, LCF/BQ/PR21/11840013. Views and opinions expressed are, however, those of the author(s) only and do not necessarily reflect those of the European Union, European Commission, European Climate, Infrastructure and Environment Executive Agency (CINEA), nor any other granting authority.  Neither the European Union nor any granting authority can be held responsible for them. 
\end{acknowledgments}

\bibliography{Primary_manuscript}





\end{document}